\documentclass[a4paper,10pt]{article}

\usepackage[english]{babel}
\usepackage[utf8]{inputenc}
\usepackage{amsmath}
\usepackage{graphicx}
\usepackage{caption}
\usepackage{subcaption}
\usepackage[a4paper,left=2.5cm,right=2.5cm,top=2.5cm,bottom=2.5cm]{geometry}
\usepackage{indentfirst}
\usepackage{authblk}
\usepackage{setspace}
\graphicspath{{figures/}}

\title{Tailoring the nucleation of domain walls along multi-segmented cylindrical nanoelements}

\author[1]{R. F. Neumann}
\author[2]{M. Bahiana}
\author[3]{S. Allende}
\author[3]{D. Altbir}
\author[4]{D. G\"orlitz}
\author[4]{K. Nielsch}

\affil[1]{IBM Research, Av. Pasteur 138 \& 146, Urca, Rio de Janeiro, CEP 22290-240, Brazil}
\affil[2]{Instituto de F\'isica, Universidade Federal do Rio de Janeiro (UFRJ), Caixa Postal 68528, Rio de Janeiro 21941-972, Brazil}
\affil[3]{Departamento de F\'isica, CEDENNA, Universidad de Santiago de Chile (USACH), Avenida Ecuador 3493, 917-0124 Santiago, Chile}
\affil[4]{Institute of Nanostructure and Solid State Physics, University of Hamburg, Jungiusstrasse 11, 20355 Hamburg, Germany}

\begin{document}
\maketitle
\onehalfspacing

\begin{abstract}
The magnetization reversal of three-segment cylindrical nanoelements comprising of alternating nanowire and nanotube sections is investigated by means of Monte Carlo simulations. Such nanoelements may feature a three-state behaviour with an intermediate plateau in the hysteresis curve due to a metastable pinning of the domain walls at the wire-tube interfaces. It turns out that vortex as well as transverse domain walls contribute to the magnetization reversal. By varying the geometric parameters, the sequence, or the material of the elements the nucleation location of domain walls, as well as their nucleation field, can be tailored.
Especially interesting is the novel possibility to drive domain walls coherently in the same or in opposite directions by changing the geometry of the hybrid nanoelement.
This important feature adds additional flexibility to the construction of logical devices based on domain wall movement.
Another prominent outcome is that domain walls can be nucleated near the centre of the element and then traverse to the outer tips of the cylindrical structure when the applied field is increased, which also opens the possibility to use these three-segment nanoelements for the field induced delivery of domain walls as substitutes for large nucleation pads.
\end{abstract}

\section{Introduction}

During the last decades much attention has been given to the development of novel nanostructures with potential for future applications.
In particular, patterned magnetic structures such as arrays of wires, tubes, dots and rings have been prepared and proposed as data storage media, spintronic devices, and even for biomedical applications~\cite{skomski2003nanomagnetics,sharma2013magnetic}.
Domain walls (DWs) in ferromagnetic wires is a subject of intense research, focusing on the search for methods that allow the injection of reproducible DWs, with well defined properties.
One of the proposal considers the use of large end pads linked to the wire~\cite{shigeto1999injection}.
When a pad is built at one end of a magnetic wire, a magnetic domain wall is always injected from the pad into the wire, making it possible to specify the direction of wall propagation in the magnetization reversal process.
On the same line, Cowburn {\it et al.}~\cite{cowburn2002domain} claim that, when using such pads, once the DW is injected, it is free to move along the wire.
Another similar experiment was presented by D. McGrouther {\it et al.}~\cite{mcgrouther2007controlled}.
In this case, the use of micron-scale pads favours the formation of a vortex DW whose chirality can be selected.

A different proposal~\cite{franken2011precise} considers the use of a He or Ga focused ion beam irradiation for controlled injection of DWs in a Pt/Co/Pt strip.
DWs experience pinning at the irradiation boundary, due to a different anisotropy at both sides of the boundary.
In the same direction, Lavrijsen {\it et al.}~\cite{lavrijsen2010controlled} used Ga$^{+}$ irradiation of Pt/Co/Pt strips to locally change the perpendicular magnetic anisotropy.
This allows the controlled injection of DWs into a device at a tunable field.

Regardless of the domain wall injection mechanism, their movement is also a field of intense research~\cite{wieser2010current,wu2014modeling}.
In a work by Wieser {\it et al.}~\cite{wieser2010current}, the current-driven motion of domain walls in ferromagnetic, cylindrical nanowires is investigated by solving the Landau-Lifshitz-Gilbert equation and the dynamic properties of transverse and vortex domain walls are extracted.
More recently, Wu {\it et al.}~\cite{wu2014modeling} studied the current-driven motion of multiple domain walls in cylindrical permalloy nanowires and investigated important factors affecting their movement.

Despite the fact that all these proposals consider the propagation of DWs in magnetic wires, a different type of nanostructure with cylindrical symmetry -- tubes -- have begun to draw much attention in the latest years.
Tubes are softer than nanowires, and offer an additional degree of freedom: the thickness of their walls.
This advantage opens new technological possibilities that have contributed to foster the research in nanotubes.

Regarding the magnetization reversal process, long nanotubes exhibit reversal modes mainly driven by two distinct mechanisms, transverse and vortex domain walls~\cite{LanderosAPL2007,proenca2012crossover}.
Both modes, according to the geometrical parameters of the tubes, show different coercivity behaviour~\cite{proenca2012crossover,escrigPRB2008,AllendeEPL2008}.
Due to the interesting physics that tubes offer, some other similar structures that include tubes have been studied.
As examples multilayer microwires have been introduced by Pirota {\it et al.}~\cite{pirota2004multilayer}.
Such structures consist of two metallic layers separated by an intermediate insulating microlayer.
In the same direction, it have been recently proposed hybrid elements composed by wires and tubes~\cite{neumann2013domain,salazar2013magnetic}, looking for the possibility of controlling the coercivity by generating a partial pinning of the domain wall at the interface between tube and wire sections. In a previous paper, Neumann \textit{et al.}~\cite{neumann2013domain} studied the possibility of controlling the domain wall propagation in order to obtain a three-state magnetic behaviour.
Such behaviour is characterized by a step or plateau in the hysteresis curve.
Focusing on the same structures, Salazar-Aravena \textit{et al.}~\cite{salazar2013magnetic} reported systematic changes in their coercivity and remanence as a function of geometry, suggesting the possibility of using them to perform logic functions~\cite{allwood2005magnetic,parkin2008magnetic}.
Such wire-tube hybrid nanoelements have, in fact, been realized experimentally by Arshad \textit{et al.}~\cite{shahid2014angular} and the change in the angular dependence of the coercivity with the length of the wire segment was studied.

Although the use of planar nanowires has been preferred in domain wall logics~\cite{allwood2005magnetic,boulle2011current,malinowski2011current}, there is a great potential for application of such hybrid cylindrical nanostructures on the construction of logical devices~\cite{allwood2005magnetic,parkin2008magnetic}.
Typically, the domain walls are driven either in opposite directions by an external magnetic field or in the same direction by an applied current~\cite{vogel2011field}.
The additional flexibility added by allowing control of the direction of motion in field-induced domain wall propagation can lead to great technological implication in the future.

\begin{figure*}[!ht]
\centering
\includegraphics[width=0.5\textwidth]{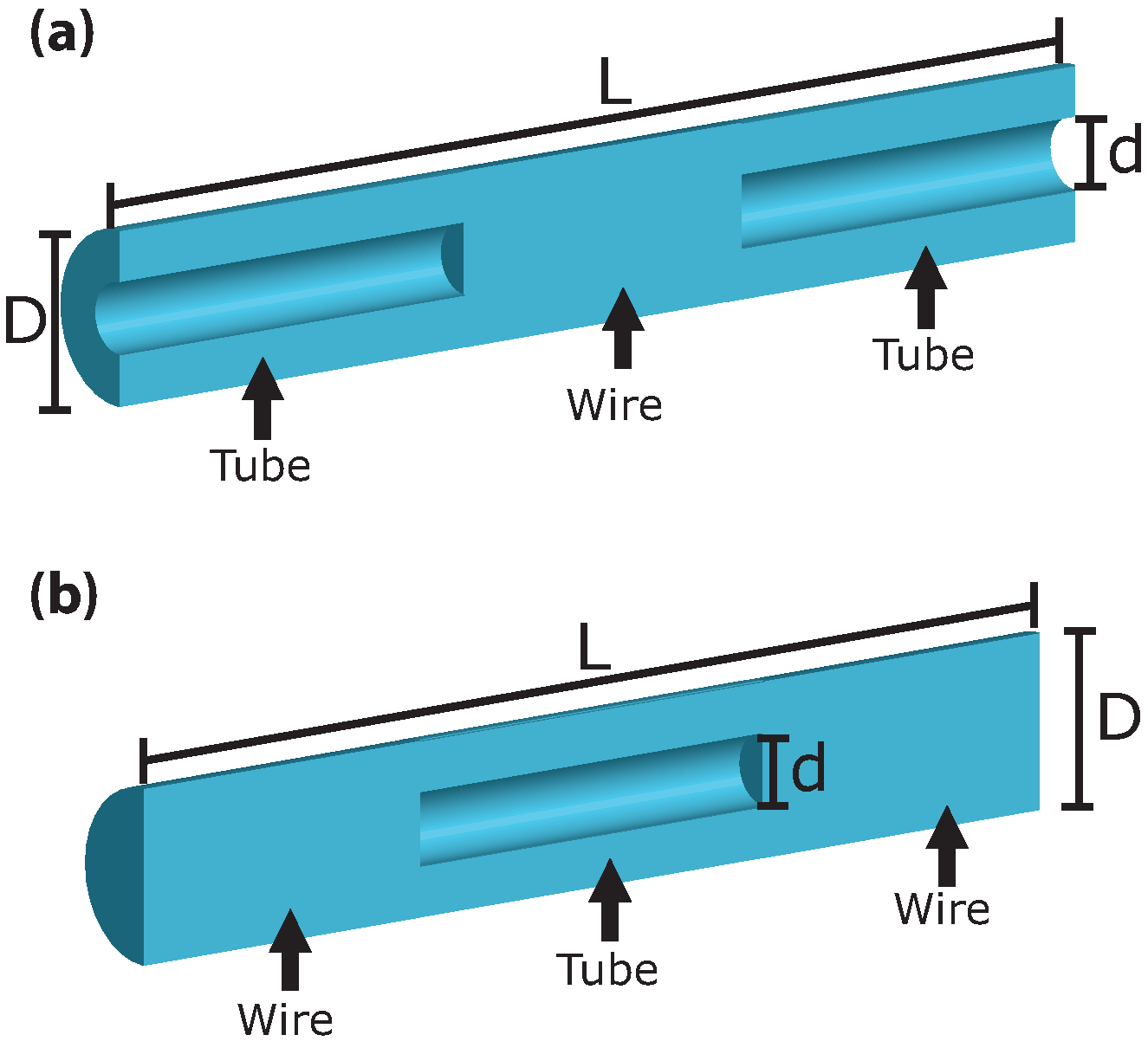}
\caption{Illustration of the (a) TWT and (b) WTW nanoelements.}
\label{fig:drawing}
\end{figure*}

Following these ideas, we explore structures composed by three elements in this article by means of Monte Carlo simulations looking for the possibility of controlling DW injection and displacement in order to tailor a multi-state magnetic behaviour.
Such property, characterized by steps or plateaus in the hysteresis curve, appears when the nanoelement is composed of sections with different coercivities, a situation that can be achieved in different ways.
Here we choose to explore the different nucleation fields associated with vortex and transverse reversal modes, the changes in the length of the structure, and the choice of materials.
Specifically, we consider  elements composed by one or two nanowires (W) and one or two nanotubes (T), defining a three-segment element, as shown in Figure~\ref{fig:drawing}.
These elements, as compared to the two-segment structures reported previously, allow to explore in detail the magnetization reversal process, separating the effect of the position (centre or tips) from other effects imposed by the geometry.

These elements, which will be called ``TWT'' or ``WTW'' from now on, can be prepared by the deposition of magnetic materials in an AAO template using a complex pulsed electrodeposition technique as a continuation and development of synthesis routes known in the literature~\cite{shahid2014angular, Garcia2014, Sada2014}.
As will be shown, and via the choice of external and internal diameters, it is possible to define, for a given material -- Ni or Co in our case --, which type of reversal mode will be preferred in the wire and in the tube sections.
This way one can engineer softer tubes and harder wires in order to obtain the desired coercivity difference leading to steps or plateaus.
This difference may be increased by the choice of different materials for the wire and tube sections.
In addition, these elements offer the possibility of controlling the place at which domain wall nucleate, offering a new alternative for large nucleation pads.

\section{Computational Model}

The energy of a nanoelement with $N$ atoms can be written as $E=-\sum_i\vec{\mu}_i\cdot \vec{H}^{\mbox{\scriptsize eff}}_i$, with an effective field given by
\begin{equation}
\vec{H}^{\mbox{\scriptsize eff}}_i = \vec{H}+K_i\left(\hat{\mu}_i \cdot \hat{e}_i\right) \hat{e}_i +\sum_j J_{ij}\vec{\mu}_j + \frac{1}{2}\sum_j\frac{3  \left(\vec \mu_j \cdot \hat n_{ij}\right)\hat{n}_{ij}-\vec \mu_j }{r_{ij}^3}\; ,
\end{equation}
where $\vec{\mu}_i$ is the magnetic moment of atom $i$ with saturation value equal to $\mu_{\mbox{\scriptsize Ni}}=0.61\,\mu_{B}$ or $\mu_{\mbox{\scriptsize Co}}=1.72\,\mu_{B}$\cite{cullity}, depending on the atomic species.
In the effective field expression, $\vec{H}$ denotes the external applied field.
In the anisotropy field contribution, $K_i$ is the anisotropy constant and $\hat{e}_i$ is the anisotropy easy axis.
The anisotropy field contribution was included for Cobalt atoms only, therefore, $K_i = 0$ for Nickel atoms and $K_i = 34~\mu\mbox{eV/atom}$ for Cobalt.
This value is equivalent to $5 \times 10^6~\mbox{erg/cm}^3$ \cite{cullity}.
In the exchange field, $J_{ij}$ is the exchange coupling constant with nonzero values only between nearest neighbours: $J_{\mbox{\scriptsize Ni-Ni}}=3.5$ meV \cite{LanderosAPL2007,AllendeEPL2008}, $J_{\mbox{\scriptsize Co-Co}}=39.5$ meV \cite{scaling1} and $J_{\mbox{\scriptsize Ni-Co}}=\sqrt{J_{\mbox{\scriptsize Ni}}J_{\mbox{\scriptsize Co}}}$ at the Ni-Co interface.
Finally, in the dipolar field term $r_{ij}$ is the distance between the $i$-th and $j$-th atoms and $\hat{n}_{ij}$ is the unitary vector along $\vec{r}_{ij}$. For simplicity we have placed both nickel and cobalt atoms in a FCC lattice with parameter $a=3.52~\mbox{\AA}$.

Dimensions of the multi-segmented nanoelements we studied in this paper are shown in Table~\ref{tab:parameters}.
Each segment is $400~\mbox{nm}$ long, so that all elements (except for WTTW4 and WTWW4) have length $L = 1200~\mbox{nm}$.
In case the of WTTW4 (WTWW4), the tube (wire) section is twice as long, leading to a total length of $1600~\mbox{nm}$.
Whenever a simple wire or tube is used for comparison, it is implicit that it has the same overall dimensions ($D$, $d$ and $L$) as the TWT or WTW that it is being compared to.

\begin{table}
\centering
\begin{tabular}{cccccc}
\hline\hline
              &  $D$ (nm) &  $d$ (nm) & $L$ (nm) &  materials   & reversal modes \\ \hline
TWT1, WTW1    &  50       &   30      & 1200     &  W-Ni  T-Ni  &  W-v  T-v      \\
TWT2, WTW2    &  30       &   15      & 1200     &  W-Ni  T-Ni  &  W-t  T-v      \\
TWT3, WTW3    &  50       &   30      & 1200     &  W-Co  T-Ni  &  W-v  T-v      \\
TWT4, WTW4    &  30       &   15      & 1200     &  W-Co  T-Ni  &  W-t  T-v      \\
WTTW4, WTWW4  &  30       &   15      & 1600     &  W-Co  T-Ni  &  W-t  T-v      \\
\hline\hline
\end{tabular}
\caption{Geometric parameters, materials and reversal modes for the nanoelements used in the simulations. In the fifth column ``v'' stands for ``vortex domain wall'' and ``t'', for ``transverse domain wall''.}
\label{tab:parameters}
\end{table}

The magnetic systems considered herein, although nanometric in scale, comprise a total of $N\approx10^8$ atoms, which is out of reach for a Monte Carlo simulation taking into account dipolar interactions with $\mathcal{O}(N^2)$ complexity.
A scaling technique \cite{scaling1} was used to reduce the number of interacting atoms that were needed correctly reproduce the magnetic behaviour of the systems.
The values of the scaling parameters are the same as in Reference~\cite{neumann2013domain}.
Based on our current computational facilities, we have designed the downscaled systems to have $6000<N^\prime<10000$ atoms.
For the scaling of the temperature we used the expression $T^\prime=\chi^{3\eta}T$~\cite{BahianaPRB2006,AllendeJAP2008}.

Following the regular Metropolis algorithm for Monte Carlo simulations, the new orientation of a magnetic moment was randomly chosen on the surface of a sphere and this change was accepted with probability $p=\mbox{min}[1,\exp (-\Delta E/k_{\mbox{\scriptsize B}} T)]$, where $\Delta E$ is the change in energy due to the reorientation of the magnetic moment and $k_B$ is the Boltzmann constant.
For out-of-equilibrium Monte Carlo simulations, such as hysteresis curves, the rate at which the parameters change is an important aspect to allow for comparison between results.
We kept the external field sweeping rate at 50 Oe per 800 Monte Carlo steps for all simulations.
This rate was chosen so that the coercive field is independent of the field sweeping rate for all systems considered.

\section{Results and Discussion}

For each of the geometries in Table~\ref{tab:parameters}, we simulated hysteresis curves at $T=300$ K with the field applied along the element axis.
Additionally, we present the differential susceptibility, $\mathrm{d}M_z/\mathrm{d}H$, as insets in the hysteresis graphs.
For each nanoelement we highlight the important points along one of the field sweeps (increasing or decreasing) of its hysteresis curve.
The points are numbered (in lowercase Roman numerals) and emphasized by arrows in the main hysteresis graph and by points overlaid to the differential susceptibility plot in the corresponding inset graph.
The snapshots of the magnetization profile corresponding to these points are also shown.
The magnetization profile is shown in terms of the normalized axial ($M_z$) and planar ($M_p$) components, where the dotted (gray) lines representing $M_i = \{0, \pm 1\}$ are guides to the eyes.
In the following we will correlate these 3 quantities -- hysteresis curve, differential susceptibility and magnetization profiles -- in order to explain the observed phenomena.

\subsection{Ni nanoelements featuring vortex domain walls (TWT1 and WTW1)}

\begin{figure*}[!ht]
\centering
\begin{subfigure}[t]{0.5\textwidth}
    \includegraphics[width=\textwidth]{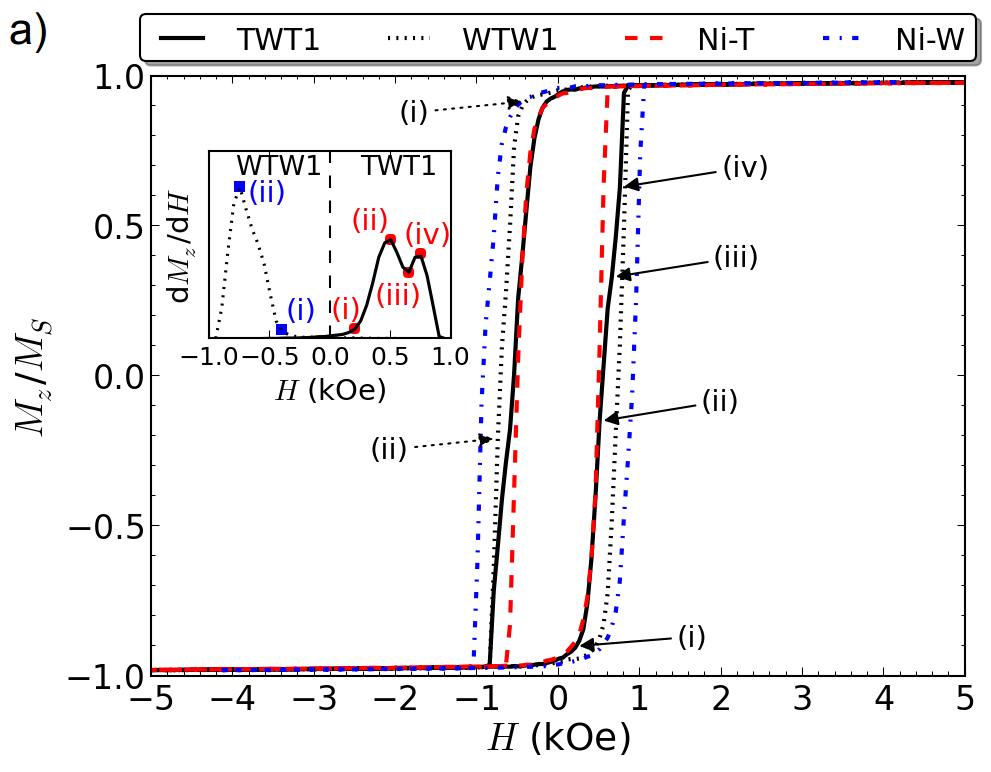}
    \caption{Hysteresis curves}
    \label{fig:TWT1-WTW1_hyst}
\end{subfigure}
\begin{subfigure}[t]{0.5\textwidth}
    \includegraphics[width=\textwidth]{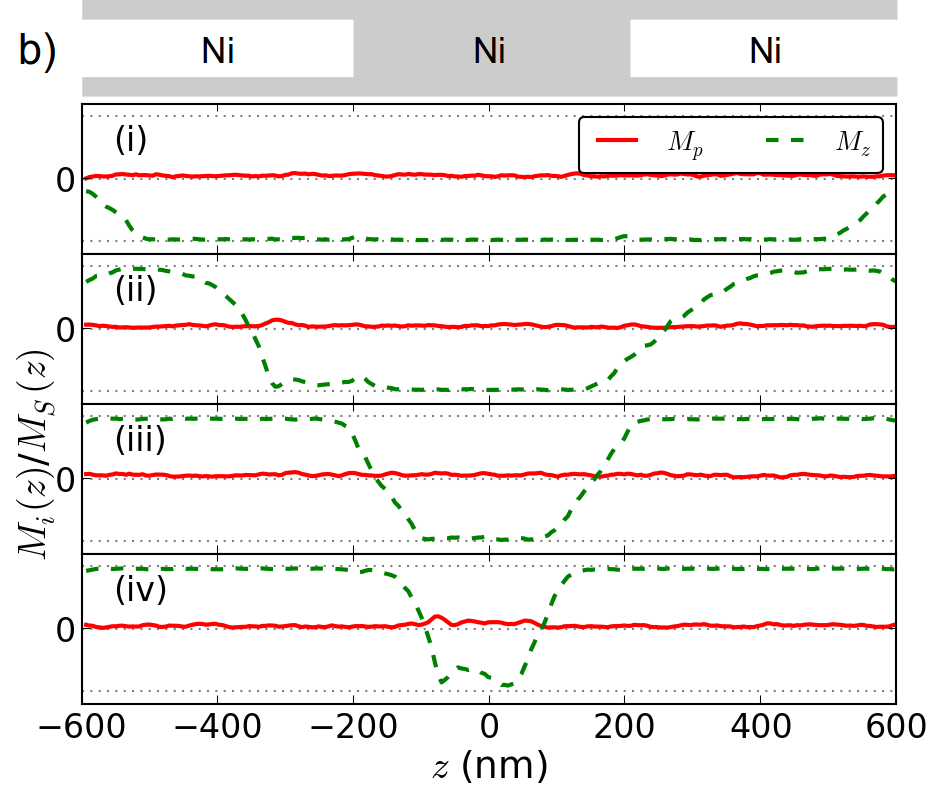}
    \caption{Magnetization profile for TWT1}
    \label{fig:TWT1_prof}
\end{subfigure}
\begin{subfigure}[t]{0.5\textwidth}
    \includegraphics[width=\textwidth]{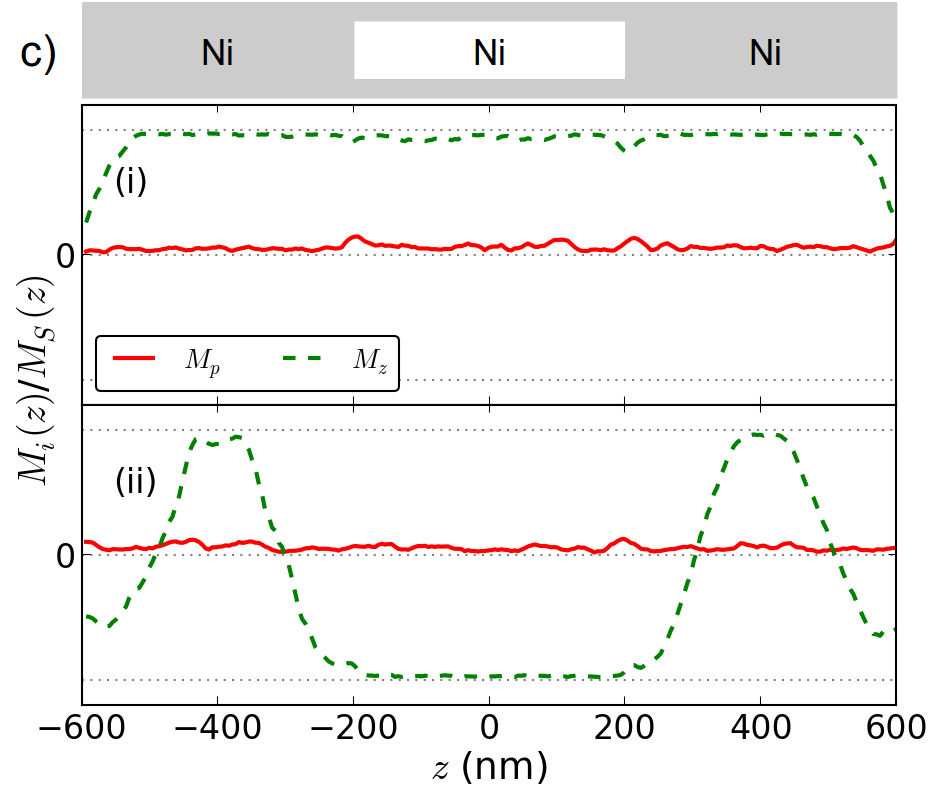}
    \caption{Magnetization profile for WTW1}
    \label{fig:WTW1_prof}
\end{subfigure}
\caption{Reversal process for elements TWT1 and WTW1 at $T = 300~\mbox{K}$, alongside with those of simple Ni tubes and wires with similar dimensions (see Table~\ref{tab:parameters}).
(\subref{fig:TWT1-WTW1_hyst}) Hysteresis curves and differential susceptibility (inset) for an applied field along the axis.
The red dots (blue squares) mark the position of the points highlighted by solid (dotted)  arrows in the main graph.
Magnetization profile for selected points during the hysteresis of (\subref{fig:TWT1_prof}) TWT1 and (\subref{fig:WTW1_prof}) WTW1.}
\label{fig:elements_1}
\end{figure*}

In Figure~\ref{fig:elements_1} we see the reversal process for elements TWT1 and WTW1 compared to those of simple Ni tubes and wires with the same overall dimensions.
The hysteresis plot itself, in Figure~\ref{fig:TWT1-WTW1_hyst}, tells very little about inner details of the reversal process but the inset graph, allied to the magnetization profiles in Figures~\ref{fig:TWT1_prof} and~\ref{fig:WTW1_prof}, allows us to observe the reversal process in detail.

The reversal of TWT1 (see Figure~\ref{fig:TWT1_prof}) starts with the emergence of vortex domain walls at the free tips (i), which then move along the tube segments towards the centre (ii) until they are partially pinned at the wire-tube interface (iii).
This partial pinning is associated with the local minimum present in Figure~\ref{fig:TWT1-WTW1_hyst} (inset).
Then, as the vortex domain walls move past the interface and the wire segment in the middle reverts its magnetization (iv) the susceptibility increases again.
The area of peaks (ii) and (iv) in the susceptibility is proportional to the magnetic volumes of the regions involved in the magnetization reversal.
Since the magnetic volume is the saturated magnetic moment, the relationship between the area below $\mathrm{d}M_z/\mathrm{d}H$ and the magnetic moment is straightforward.

While the nucleation field for TWT1 remained equal to that of a simple Ni tube, a larger value (closer to that of a simple Ni wire) was observed for WTW1 due to the fact that the (softer) Ni tube segment is in the middle of the nanoelement, thus, far from edge effects that usually help the nucleation of domain walls.
This increases the similarity between the nucleation fields for the tube and wire segments which, in turn, leads to an apparently trivial reversion process characterized by a single peak in the differential susceptibility as depicted in Figure~\ref{fig:TWT1-WTW1_hyst} (inset).
In Figure~\ref{fig:WTW1_prof}, however, we observe a non-trivial reversal process that starts with the nucleation of vortex domain walls at the free tips of the wire segments (i) but, before these domain walls manage to cross the wire segments, the tube segment in the middle reverts as a whole (ii), leading to the observed peak in the susceptibility.
The presence of the wire segments at the extremities of the central tube segment induce a different non-trivial reversal mode due the absence of an obvious nucleation site.

\subsection{Ni nanoelements featuring vortex and transverse domain walls (TWT2 and WTW2)}

\begin{figure*}[!ht]
\centering
\begin{subfigure}[t]{0.5\textwidth}
    \includegraphics[width=\textwidth]{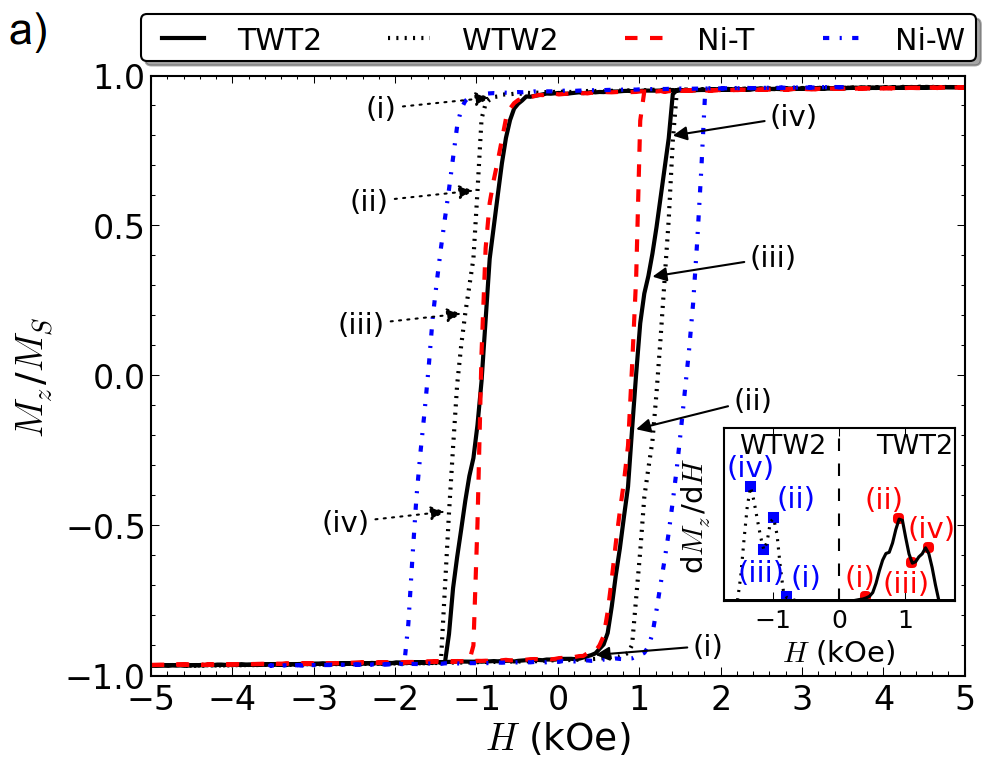}
    \caption{Hysteresis curves}
    \label{fig:TWT2-WTW2_hyst}
\end{subfigure}
\begin{subfigure}[t]{0.5\textwidth}
    \includegraphics[width=\textwidth]{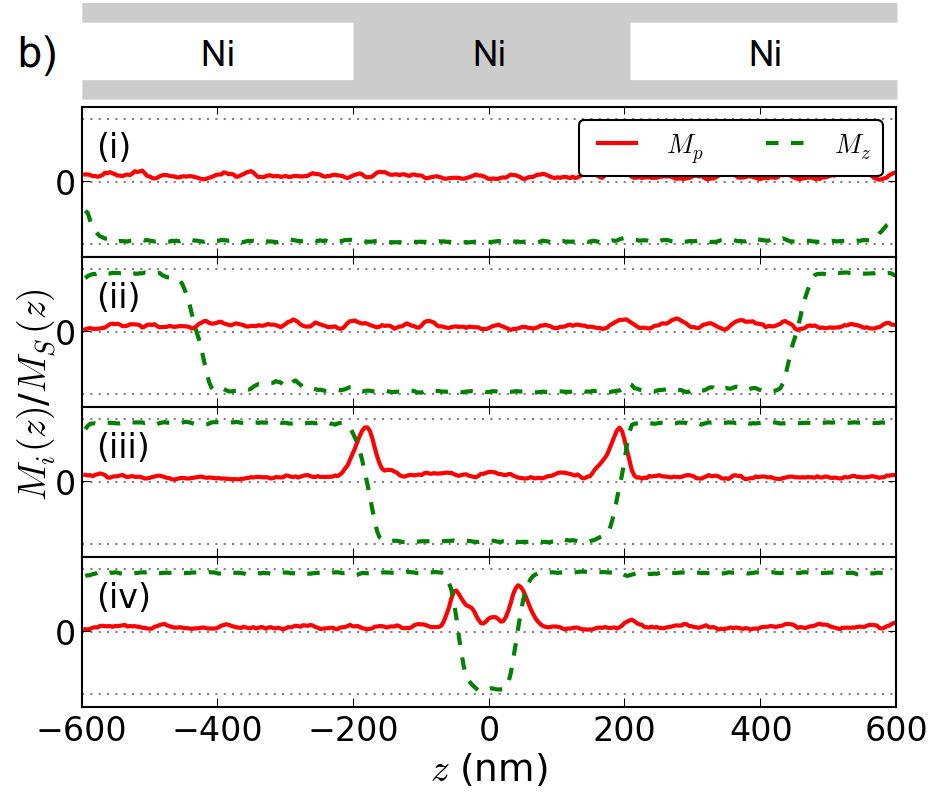}
    \caption{Magnetization profile for TWT2}
    \label{fig:TWT2_prof}
\end{subfigure}
\begin{subfigure}[t]{0.5\textwidth}
    \includegraphics[width=\textwidth]{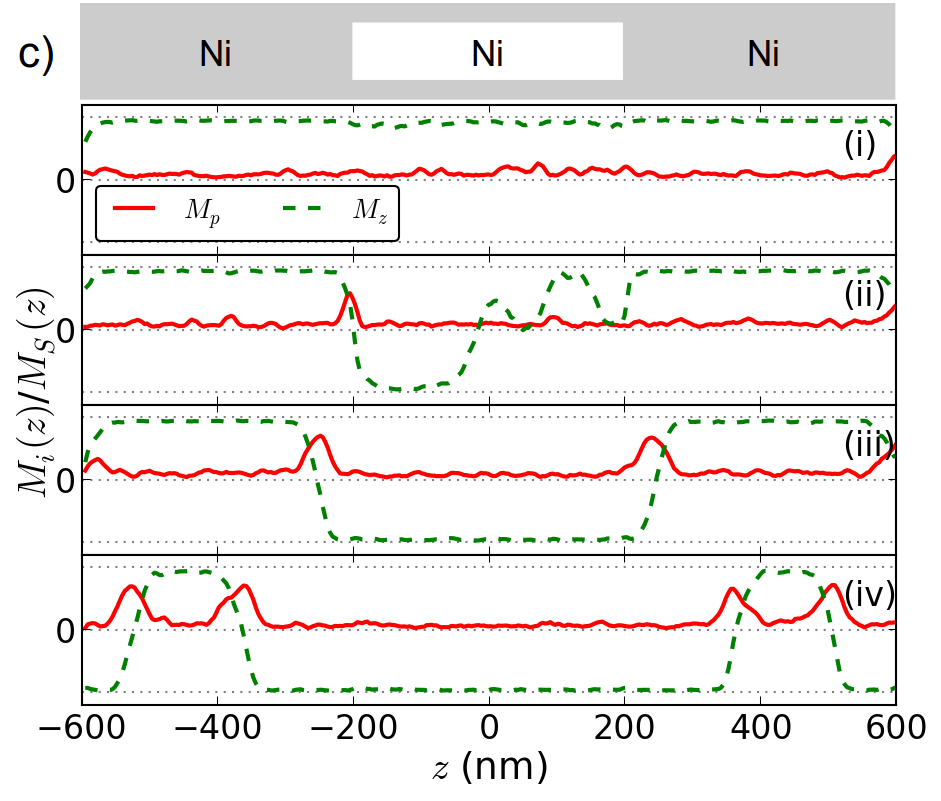}
    \caption{Magnetization profile for WTW2}
    \label{fig:WTW2_prof}
\end{subfigure}
\caption{Reversal process for elements TWT2 and WTW2 at $T = 300~\mbox{K}$, alongside with those of simple Ni tubes and wires with similar dimensions (see Table~\ref{tab:parameters}).
(\subref{fig:TWT2-WTW2_hyst}) Hysteresis curves and differential susceptibility (inset) for an applied field along the axis.
The red dots (blue squares) mark the position of the points highlighted by solid (dotted)  arrows in the main graph.
Magnetization profile for selected points during the hysteresis of (\subref{fig:TWT2_prof}) TWT2 and (\subref{fig:WTW2_prof}) WTW2.}
\label{fig:elements_2}
\end{figure*}

By changing the radial dimensions of the segments, one can induce transverse domain walls in the wire segments, which have higher nucleation fields as compared to vortex domain walls.
This increase in the difference between nucleation fields in the tube and wire segments leads to an enhancement of the pinning effect, as shown in Figure~\ref{fig:TWT2-WTW2_hyst}.
In this case, both TWT2 and WTW2 present the same double peak -- points (ii) and (iv) -- structure for the differential susceptibility with the minimum between them -- point (iii) -- representing the partial pinning of the domain wall.
The first peak (ii) is related to the reversal of (softer) Ni tube segments and the second peak (iv), to the reversal of (harder) Ni wire segments.

In Figure~\ref{fig:TWT2_prof}, we show the reversal process for TWT2 in terms of its magnetization profiles at selected points during the hysteresis.
The reversal starts with the formation of vortex domain walls at the tips of the outer Ni tube segments (i), that later move towards the central wire section (ii) until they are partially pinned at the interface (iii).
At this point, a transversal domain wall is formed at the interface.
Finally, the newly created transversal domain walls meet at the centre of the wire section (iv).

The magnetization profiles during the hysteresis of WTW2 are depicted in Figure~\ref{fig:WTW2_prof}.
When the nucleation of transversal domain walls at the tips of the outer wires has barely started (i), the central tube section reverts showing the absence of a specific site for nucleation (ii).
Then, as the domain walls reach the interface and are partially pinned (iii) their transverse character is strengthened.
Finally, transverse domain walls from the outer tips and moving towards the centre put an end to the reversal as they meet with the outward-moving domain walls (iv).
The meeting point of the inward- and outward-moving domain walls is clearly not the centre of the outer wire segments, which denotes a reduction of the nucleation field at the wire-tube interfaces with respect to the free wire tips.

\subsection{Nanoelements with Ni and Co segments featuring vortex domain walls (TWT3 and WTW3)}

\begin{figure*}[!ht]
\centering
\begin{subfigure}[t]{0.5\textwidth}
    \includegraphics[width=\textwidth]{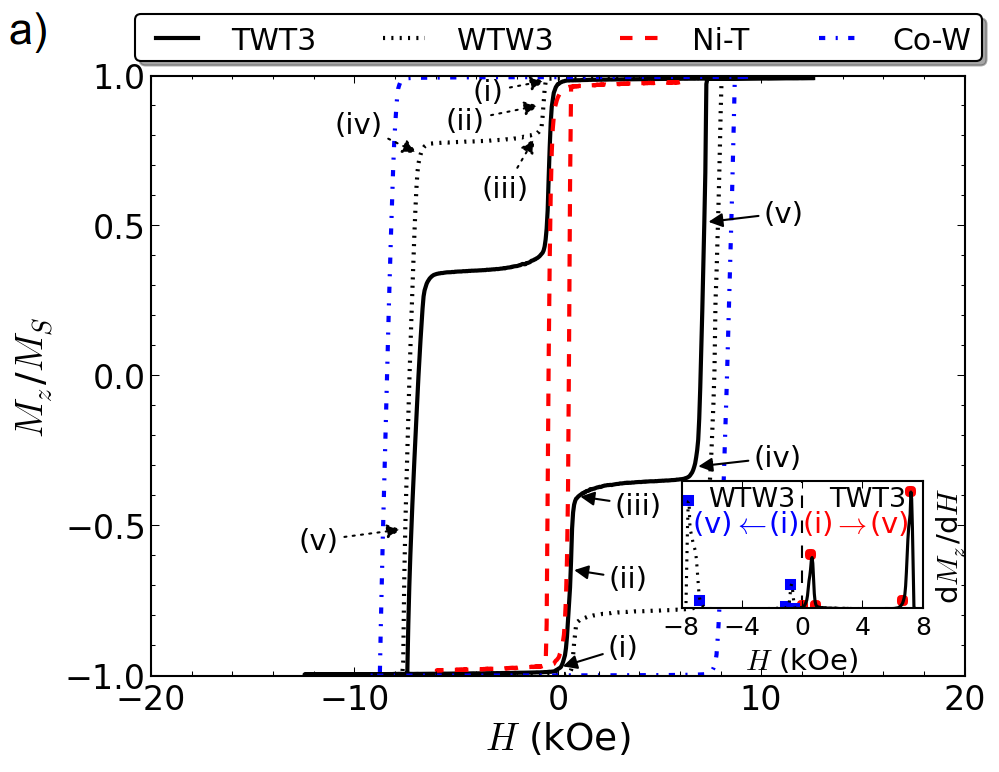}
    \caption{Hysteresis curves}
    \label{fig:TWT3-WTW3_hyst}
\end{subfigure}
\begin{subfigure}[t]{0.5\textwidth}
    \includegraphics[width=\textwidth]{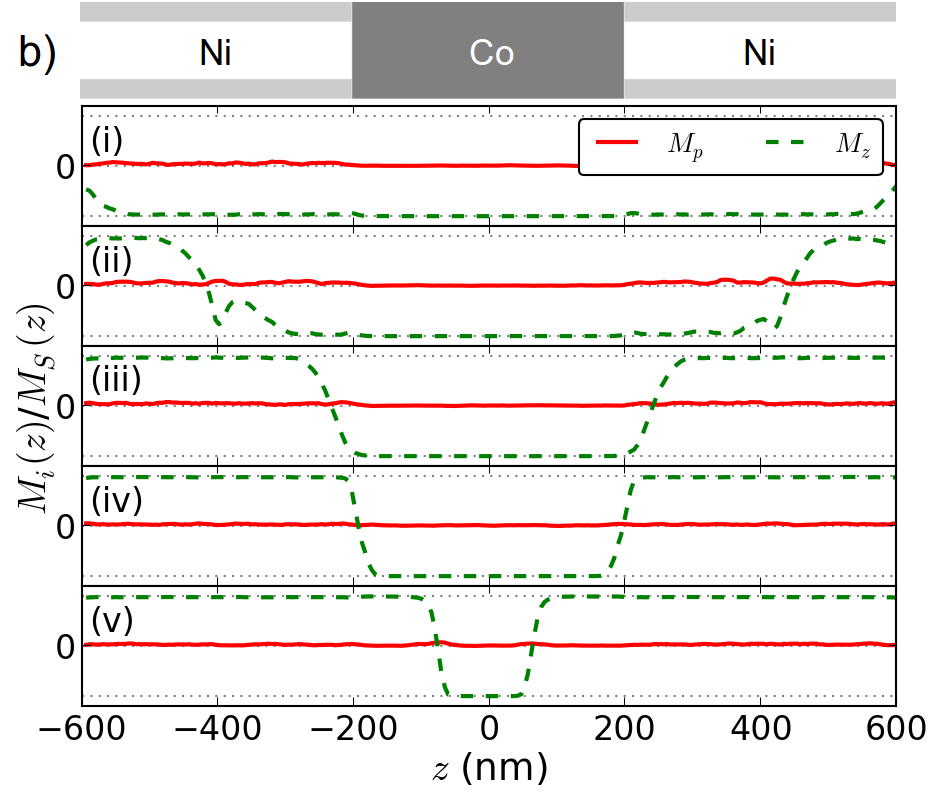}
    \caption{Magnetization profile for TWT3}
    \label{fig:TWT3_prof}
\end{subfigure}
\begin{subfigure}[t]{0.5\textwidth}
    \includegraphics[width=\textwidth]{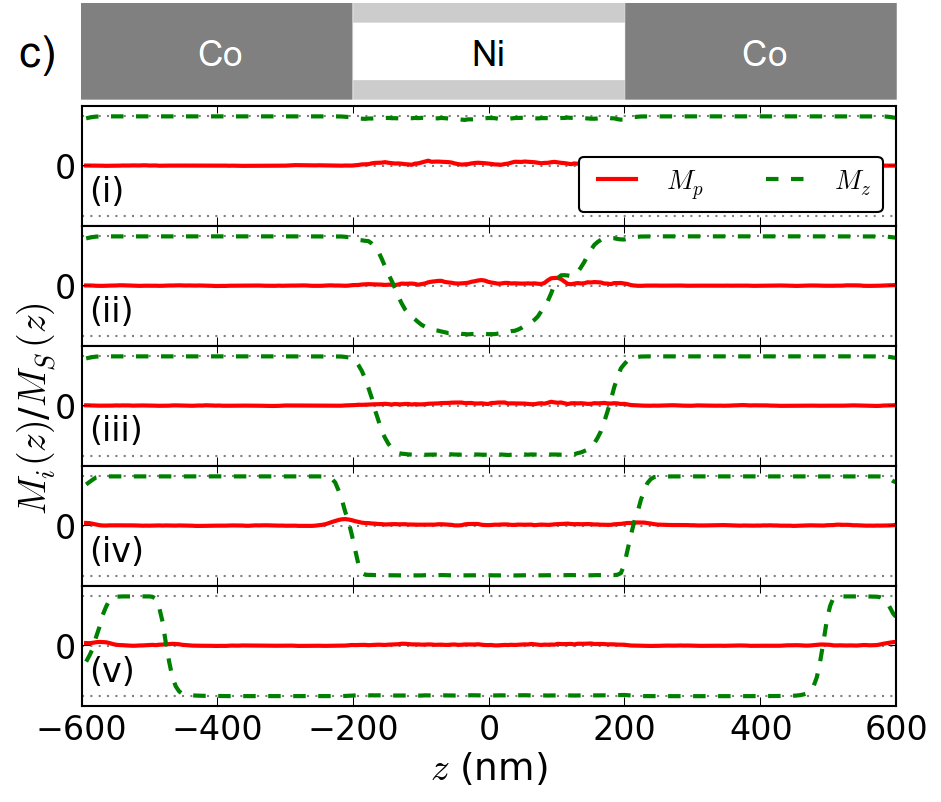}
    \caption{Magnetization profile for WTW3}
    \label{fig:WTW3_prof}
\end{subfigure}
\caption{Reversal process for elements TWT3 and WTW3 at $T = 300~\mbox{K}$, alongside with those of simple Ni tubes and Co wires with similar dimensions (see Table~\ref{tab:parameters}). (\subref{fig:TWT3-WTW3_hyst}) Hysteresis curves and differential susceptibility (inset) for an applied field along the axis.
The red dots (blue squares) mark the position of the points highlighted by solid (dotted) arrows in the main graph.
Magnetization profile for selected points during the hysteresis of (\subref{fig:TWT3_prof}) TWT3 and (\subref{fig:WTW3_prof}) WTW3.}
\label{fig:elements_3}
\end{figure*}

An extra source of difference between wire and tube segments was added by making the wire segments even harder with the switch from Ni to Co as the constituting species.
The reversal modes in each segment are the same as in TWT1 and WTW1 due to the choice of diameters.
The huge difference in nucleation fields generates a plateau having a width of $\approx 5.8~\mbox{kOe}$ in the hysteresis curve, as noted previously~\cite{neumann2013domain,salazar2013magnetic}.
The differential susceptibility presents (see inset in Figure~\ref{fig:TWT3-WTW3_hyst}), therefore, two peaks separated by a null region.
The first peak (ii) is related to the reversal of the (softer) Ni tube segment and the second (v) to that of the (harder) Co wire segment.
The plateau lies between points (iii) and (iv) in Figure~\ref{fig:TWT3-WTW3_hyst}, during which the domain walls cross the wire-tube interfaces (see Figures~\ref{fig:TWT3_prof} and ~\ref{fig:WTW3_prof}).
In both cases, points (iii) and (iv) represent the passing of the head and the tail of the domain wall, respectively, at the interface.

For TWT3 (see Figure~\ref{fig:TWT3_prof}), the reversal starts at the tips of the Ni tube segments (i) and evolves until the vortex domain walls meet at the centre of the Co wire segment (v).
On the other hand, WTW3 (see Figure~\ref{fig:WTW3_prof}) starts reverting from the centre (ii) where the Ni tube segment is located.
After passing by the interface between points (iii) and (iv), the vortex domain walls move outwards along the wire segments until they meet with inward-moving domain walls coming from the tips.
It is striking, though, to note that this meeting point is quite far from the centre of the wire segment, which demonstrates the large difference in the nucleation fields at the two extremities (the free and the interface) of the outer wire segments.

\subsection{Nanoelements with Ni and Co segments featuring vortex and transverse domain walls (TWT4 and WTW4)}

\begin{figure*}[!ht]
\centering
\begin{subfigure}[t]{0.5\textwidth}
    \includegraphics[width=\textwidth]{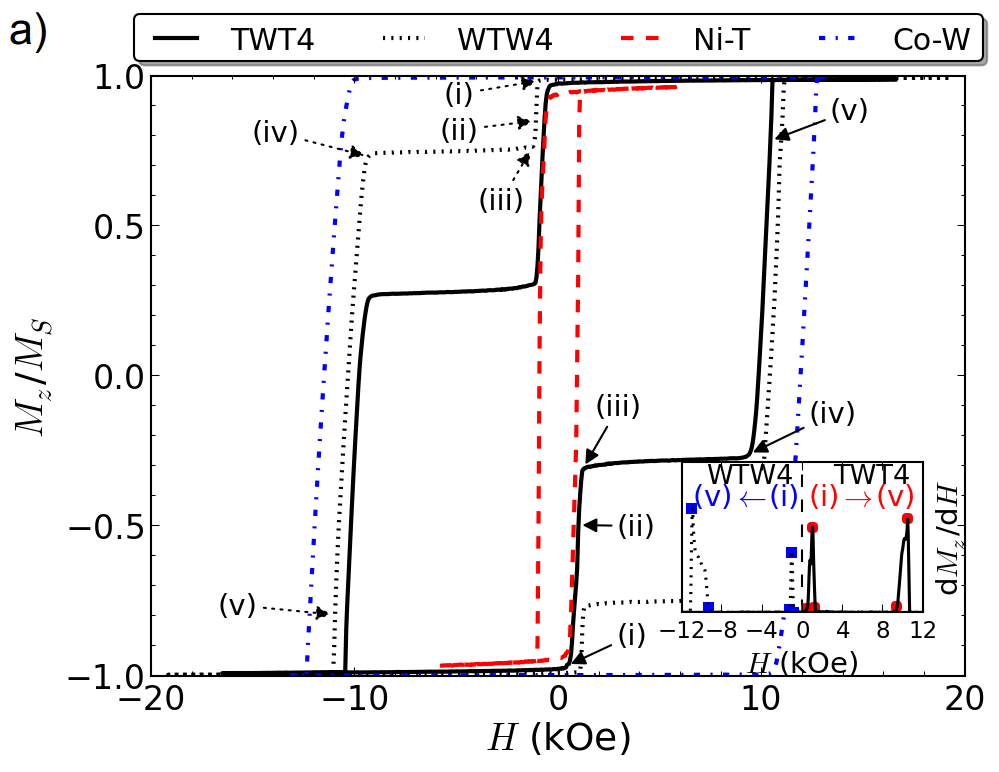}
    \caption{Hysteresis curves}
    \label{fig:TWT4-WTW4_hyst}
\end{subfigure}
\begin{subfigure}[t]{0.5\textwidth}
    \includegraphics[width=\textwidth]{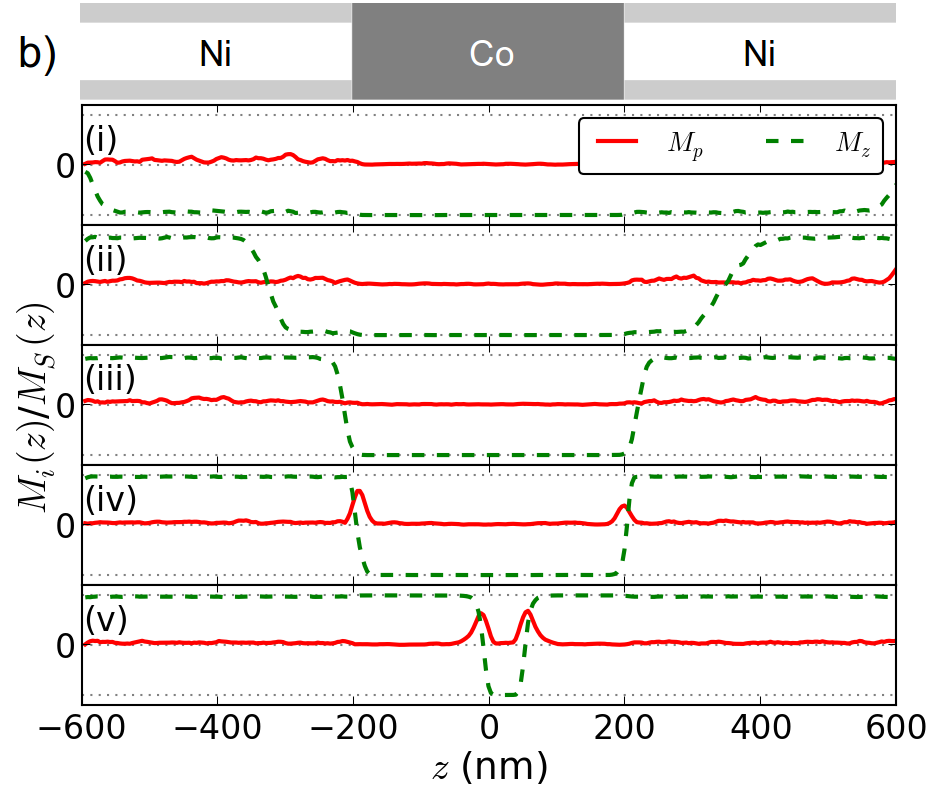}
    \caption{Magnetization profile for TWT4}
    \label{fig:TWT4_prof}
\end{subfigure}
\begin{subfigure}[t]{0.5\textwidth}
    \includegraphics[width=\textwidth]{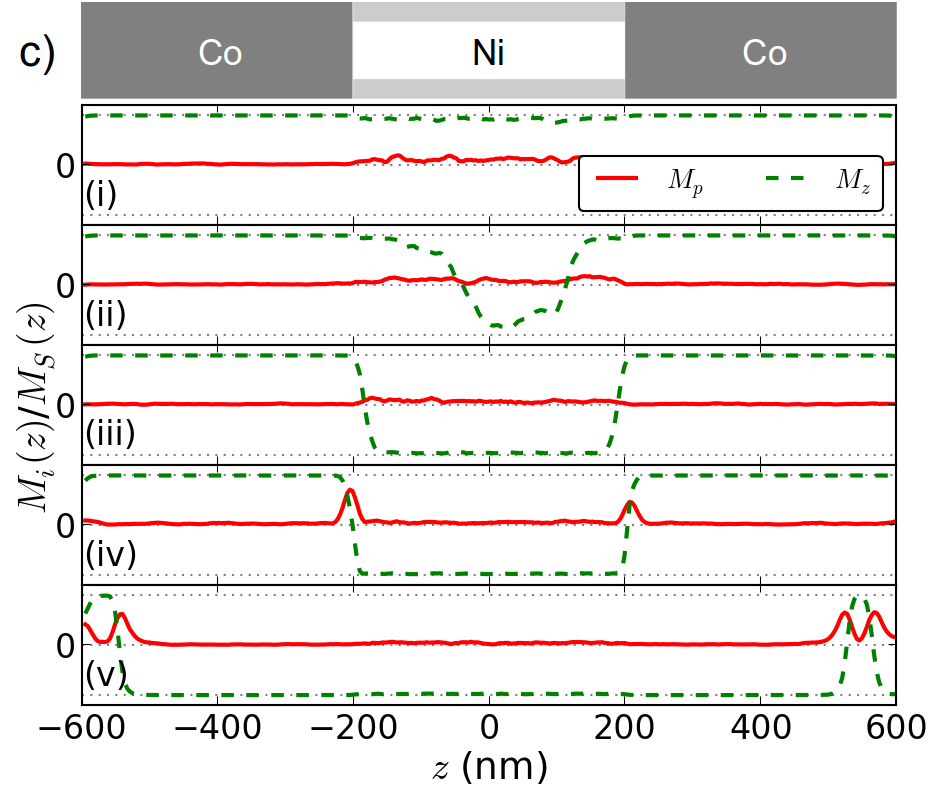}
    \caption{Magnetization profile for WTW4}
    \label{fig:WTW4_prof}
\end{subfigure}
\caption{Reversal process for elements TWT4 and WTW4 at $T = 300~\mbox{K}$, alongside with those of simple Ni tubes and Co wires with similar dimensions (see Table~\ref{tab:parameters}).
(\subref{fig:TWT4-WTW4_hyst}) Hysteresis curves and differential susceptibility (inset) for an applied field along the axis.
The red dots (blue squares) mark the position of the points highlighted by solid (dotted) arrows in the main graph.
Magnetization profile for selected points during the hysteresis of (\subref{fig:TWT4_prof}) TWT4 and (\subref{fig:WTW4_prof}) WTW4.}
\label{fig:elements_4}
\end{figure*}

By allying the choice of materials with the choice of reversal modes one can further increase the effects studied herein.
Due to the even larger difference of nucleation fields for the transversal domain wall in the Co wire and the vortex domain wall in the Ni tube, we obtain a larger plateau of, approximately, $8.1~\mbox{kOe}$ between points (iii) and (iv) in Figure~\ref{fig:elements_4}.
The two-peak structure for the differential susceptibility is maintained (see Figure~\ref{fig:TWT4-WTW4_hyst}).

As in the previous cases, for TWT4 (see Figure~\ref{fig:TWT4_prof}) the Ni tubes start reverting via vortex domain walls at the tips of the element (i), which then move freely (ii) until stopped by the interface (iii).
Later on, the domain walls at the interface change from vortex to transversal (iv) and then move towards the centre of the element (v).
WTW4, as shown in Figure~\ref{fig:WTW4_prof}, starts reverting by its central Ni tube segment (ii).
The vortex domain walls stay at the wire-tube interfaces (iii) until their transformation to transverse domain walls is completed (iv).
Finally, the transverse domain walls move along the wire segments (v), completing the reversal.
For this element, only when the outward-moving domain walls are about to reach the tips the extremities start reverting as well.
It is noteworthy that, during most of the reversal of the wire segments, there was only one domain wall present per segment, which shows the high degree to which the controlled nucleation and manipulation of domain walls at specific locations can be tailored.
It should be noted that the nanoelement WTW4 could be utilized as a domain wall delivery device at its tips when the field is inverted.

\begin{figure*}[!ht]
\centering
\begin{subfigure}[t]{0.5\textwidth}
    \includegraphics[width=\textwidth]{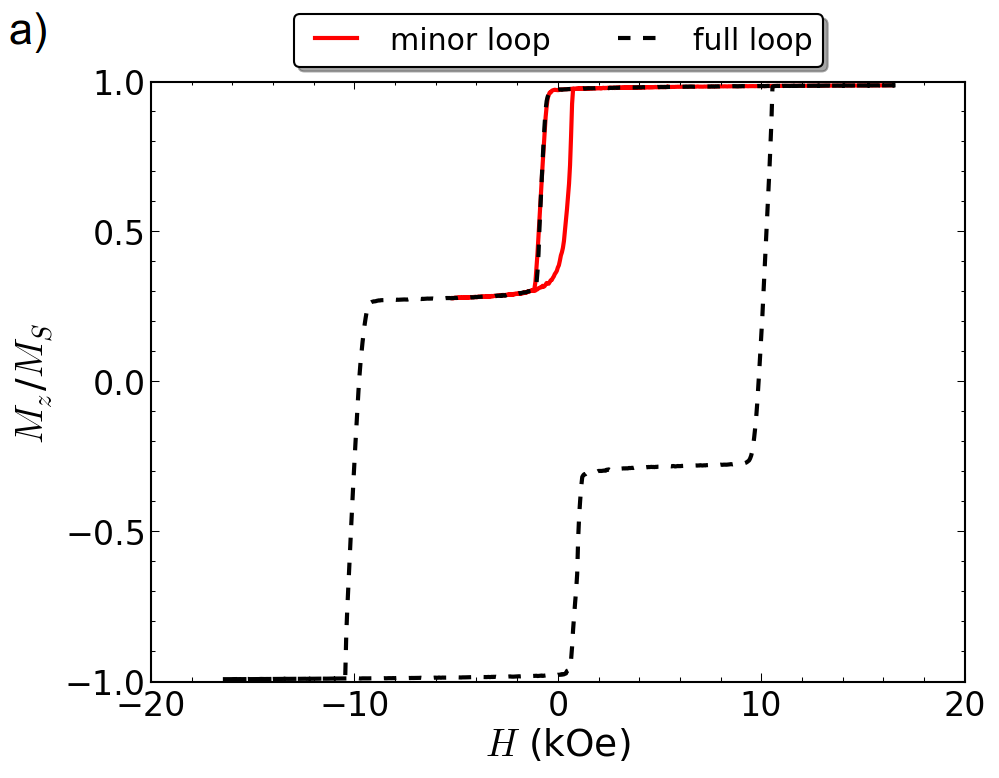}
    \caption{Minor hysteresis loop for TWT4}
    \label{fig:TWT4_minor_loop}
\end{subfigure}
\begin{subfigure}[t]{0.5\textwidth}
    \includegraphics[width=\textwidth]{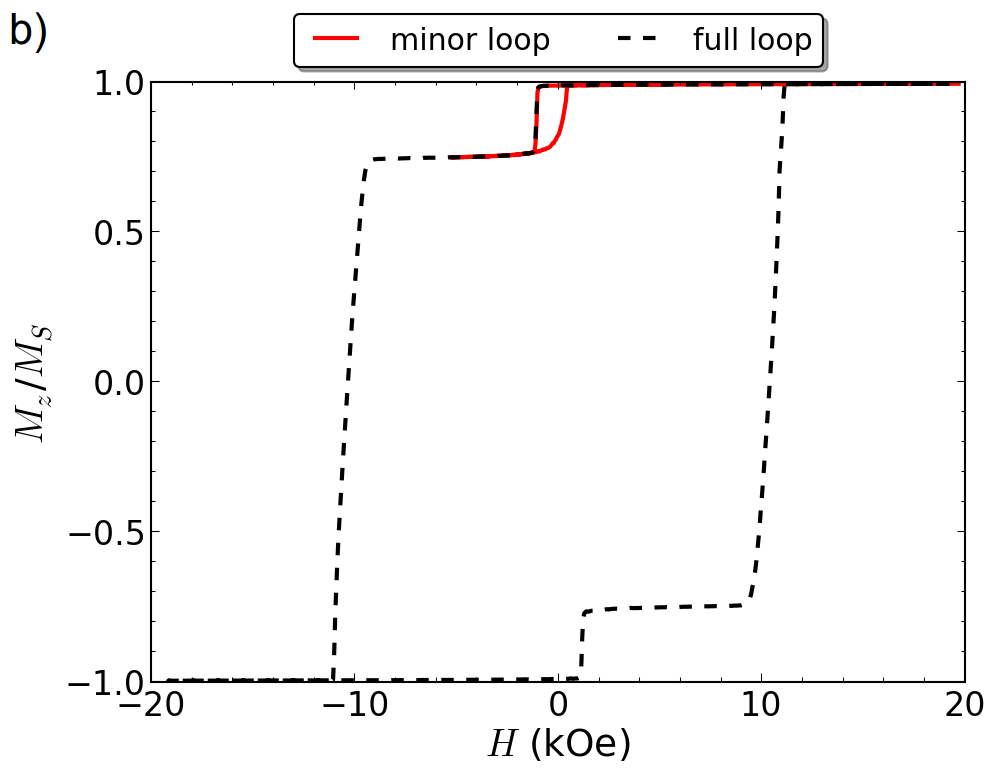}
    \caption{Minor hysteresis loop for WTW4}
    \label{fig:WTW4_minor_loop}
\end{subfigure}
\caption{Minor hysteresis loops for (\subref{fig:TWT4_minor_loop}) TWT4 and (\subref{fig:WTW4_minor_loop}) WTW4, where the applied field was swept until the middle of the plateau and then swept back in the opposite direction.}
\label{fig:minor_loop}
\end{figure*}

In order to check whether the intermediate magnetization state associated with the intermediate plateau is stable, we performed simulations of minor hysteresis loops for these nanoelements.
The procedure for the minor hysteresis loops consisted of sweeping the applied field until the middle of the plateau and then sweeping it back in the other direction.
As shown in Figure~\ref{fig:minor_loop}, such intermediate states are stable and can be easily manipulated by a low applied field.
The minor hysteresis loop of TWT4 in Figure~\ref{fig:TWT4_minor_loop} has a half-width of 670 Oe and is displaced by -190 Oe with respect to zero.
The values for WTW4 as displayed in Figure~\ref{fig:WTW4_minor_loop} are 645 Oe and -445 Oe, respectively.
Especially, Figures~\ref{fig:TWT4_minor_loop} and~\ref{fig:WTW4_minor_loop} give rise to a new technique: by magnetic field pulses used for this minor loop one can repeatedly create domain walls at the outer ends.

\subsection{Segments with unequal lengths (WTTW4 and WTWW4)}

In view of the previous subsection, in which a domain wall was injected into a harder segment from a softer one, we test the validity of our previous conclusions by considering similar systems in which one of the segments, either the tube or the wire, is twice as long as in WTW4.
Therefore, starting from WTW4, we doubled the length of the central tube (outer wire) segment, obtaining the system called WTTW4 (WTWW4) as described in Table~\ref{tab:parameters}.

\begin{figure*}[!ht]
\centering
\begin{subfigure}[t]{0.5\textwidth}
    \includegraphics[width=\textwidth]{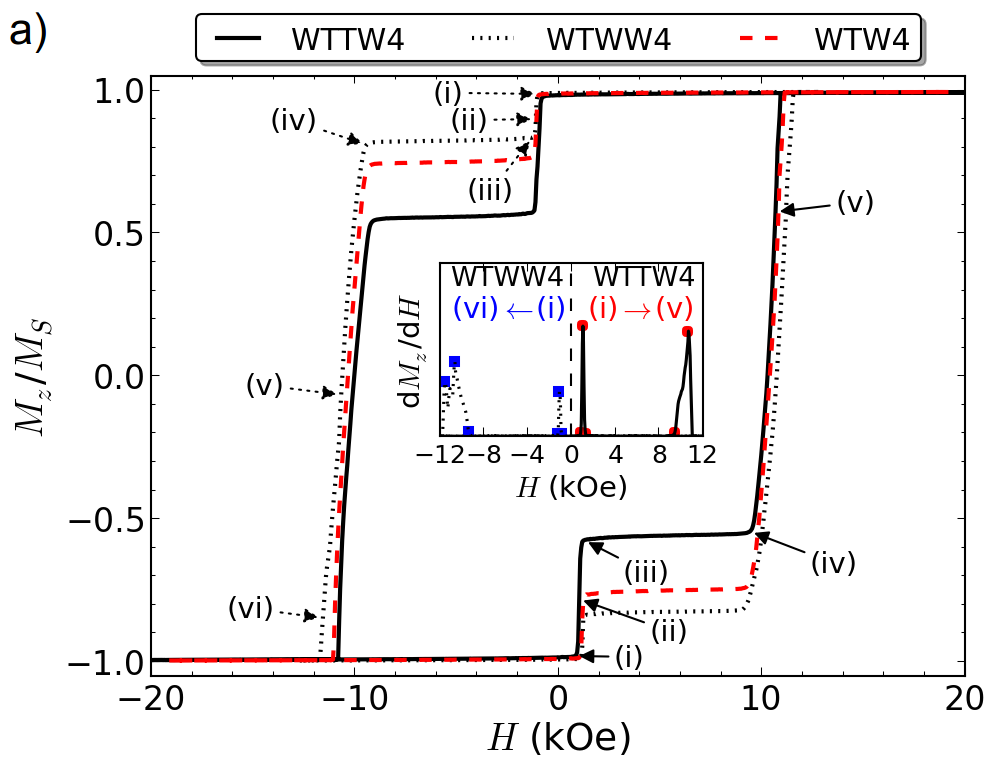}
    \caption{Hysteresis curves}
    \label{fig:WTTW4-WTWW4_hyst}
\end{subfigure}
\begin{subfigure}[t]{0.5\textwidth}
    \includegraphics[width=\textwidth]{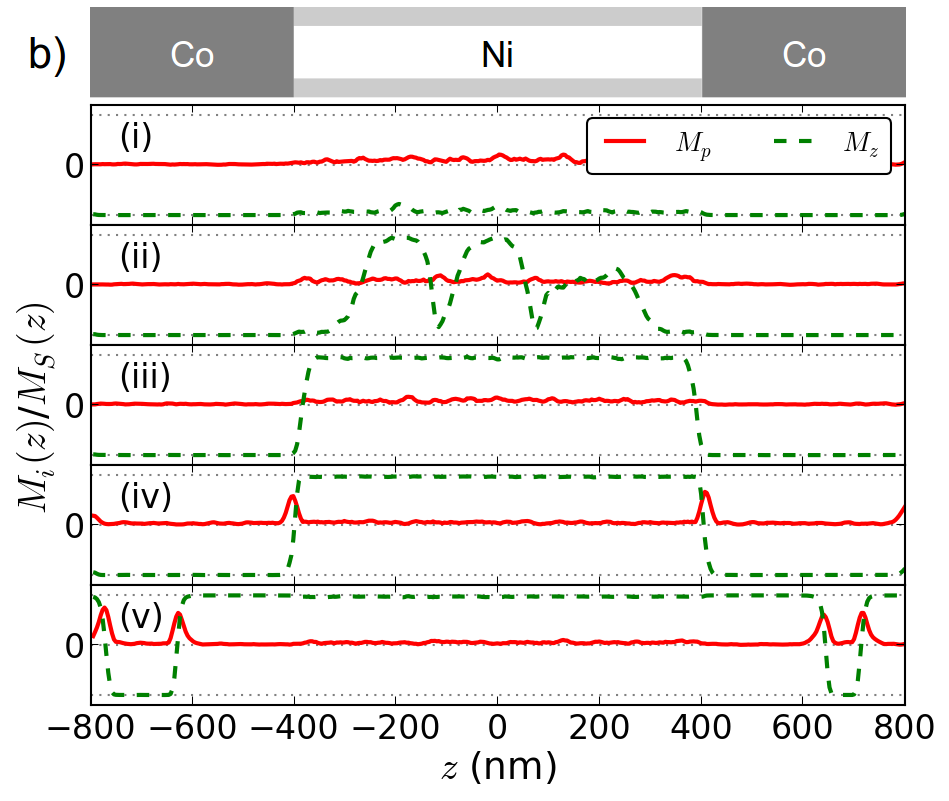}
    \caption{Magnetization profile for WTTW4}
    \label{fig:WTTW4_prof}
\end{subfigure}
\begin{subfigure}[t]{0.5\textwidth}
    \includegraphics[width=\textwidth]{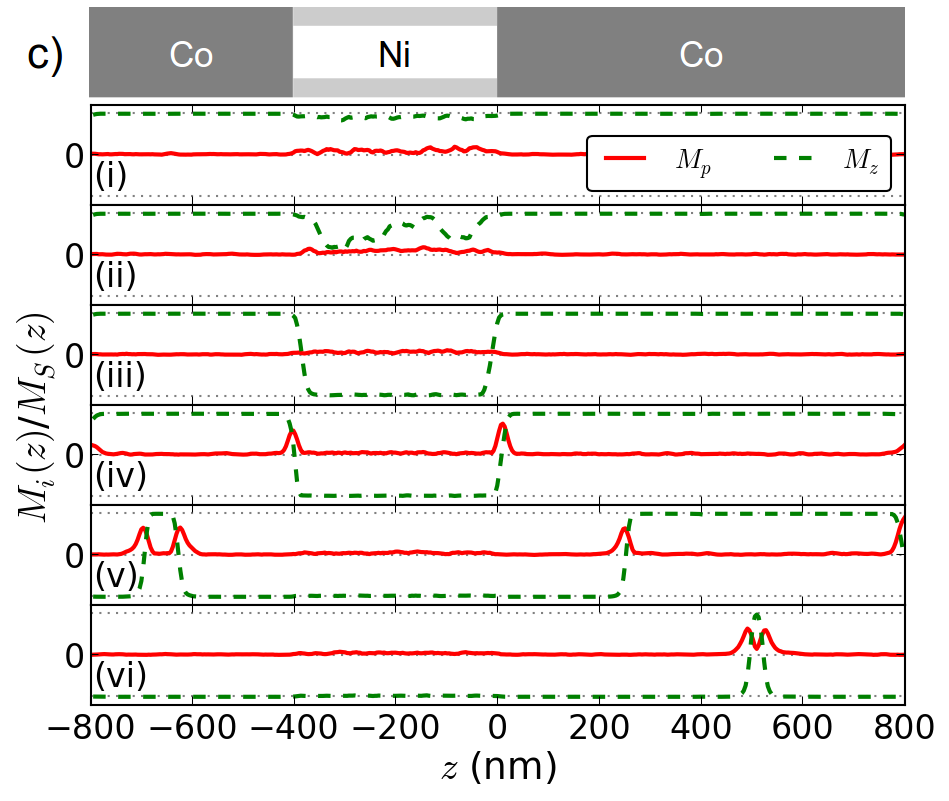}
    \caption{Magnetization profile for WTWW4}
    \label{fig:WTWW4_prof}
\end{subfigure}
\caption{Reversal process for elements WTTW4 and WTWW4 at $T = 300~\mbox{K}$, alongside with that of its parent system WTW4 (see Table~\ref{tab:parameters}).
The difference between WTTW4 (WTWW4) and WTW4 is that the central Ni tube (outer Co wire) segment is twice as large.
(\subref{fig:WTTW4-WTWW4_hyst}) Hysteresis curves and differential susceptibility (inset) for an applied field along the axis.
The red dots (blue squares) mark the position of the points highlighted by solid (dotted) arrows in the main graph.
Magnetization profile for selected points during the hysteresis of (\subref{fig:WTTW4_prof}) WTTW4 and (\subref{fig:WTWW4_prof}) WTWW4.}
\label{fig:elements_4long}
\end{figure*}

Comparing WTTW4 and WTW4 in Figure~\ref{fig:WTTW4-WTWW4_hyst}, one sees that the only change in the hysteresis curve caused by the increase in length of the central Ni tube segment with respect to WTW4 is the larger drop in magnetization after the reversal of the tube, due to the larger fraction of the magnetic volume associated to this softer segment.
The inset in this figure also shows for WTTW4 the same two-peak structure of the differential susceptibility observed for WTW4.
The similarity between those systems is corroborated by comparing Figure~\ref{fig:WTTW4_prof} to~\ref{fig:WTW4_prof} and noting the clear correspondence between points (i)--(v).

For WTWW4 on the other hand, although the hysteresis curve in Figure~\ref{fig:WTTW4-WTWW4_hyst} is very similar to that of WTW4, the presence of the longer Co wire at one end changed considerably the reversal process, evidenced by a third peak in the differential susceptibility appearing at higher fields (inset).
The magnetization profile in Figure~\ref{fig:WTWW4_prof} shows that the second (v) and third (vi) susceptibility peaks are due to the reversal of the shorter and longer Co wire segment, respectively.
The outward-moving domain walls coming from the Ni tube segment clearly move more than the inward-moving ones coming from the tips, confirming the success of the domain wall injection.

It is important, however, to note that for both WTTW4 and WTWW4, the meeting point of the inward- and outward-moving domain walls is closer to the centre of the wire segments when compared to WTW4.
This fact highlights that the length of the segments plays a key role in determining the effectiveness of the injection of domain walls.
If the external magnetic field is varied only in a small range, that is, along the fields at which appear the plateau,  one can switch repeatedly between the intermediate plateau and saturation. This is depicted in Figure~\ref{fig:minor_loop} for the nanoelements TWT4 (\subref{fig:TWT4_minor_loop}) and WTW4 (\subref{fig:WTW4_minor_loop}). This reproducible behaviour opens the possibility for using such three-segment elements for domain wall injection.

\section{Conclusions}

A systematic study of the magnetization reversal in three-segment cylindrical nanoelements comprising of alternating nanowire and nanotube segments has been performed using Monte Carlo simulations.
By choosing different sets of parameters and materials for the tube and wire segments, the nucleation of vortex and/or transverse domain walls could be tailored, thus creating a manifold of different scenarios for the magnetization reversal.
It turned out that by using different lengths for the individual segments, the location for domain wall annihilation along the element axis could be varied.

In all cases, regardless of the position of the wire and tube segments, the magnetization reversal finishes first in the tube segments, even when it starts elsewhere (see Figure~\ref{fig:WTW1_prof}).
When there is only one material (except for WTW1), the arrival of the domain wall at the interface leads to a local minimum in the differential susceptibility, since the difference in nucleation field across the interface is not large enough to create a plateau.
However, when different materials are used to harden (soften) the wire (tube) section, an intermediate plateau is formed due to the larger difference of nucleation fields on each side of the interface.
In general, for the WTW-like elements where the lower coercivity segment is at the centre, the reversal starts at this segment but, due to the lack of a specific nucleation centre, it might start anywhere in this softer section, leading to a irregular profile (see Figure~\ref{fig:WTW2_prof})
For all cases in which an intermediate plateau was observed, the width of the wall decreases notably as it reaches the interface, regardless of the reversal mode.
It is important to highlight the absence of a preferential nucleation site for domain walls in the WTW elements because in these elements the tube ends are fixed to the wires, which are harder, preventing the nucleation of domain walls at the interfaces. Moreover, because the tube is uniform, the wall can nucleate almost simultaneously in different places,  opening the possibility of injecting walls almost simultaneously on both wires by varying the size of the tube. In summary, a novel feature in cylindrical nanostructures was found, which consists of the outward movement of the domain walls from the element body to the element tips. This might give rise to applications in which such nanoelements are used as domain wall delivering structures at its tips.

\section*{Acknowledgements}
D.G. gratefully acknowledges constructive discussions with R. Zierold (University of Hamburg) and A. Espejo (USACH, on research visit to University of Hamburg). We acknowledge financial support in Chile from FONDECYT 1120356 and 11121214, Grant ICM P10-061-F by FIC-MINECON, Financiamiento Basal para Centros Cient\'ificos y Tecnol\'{o}gicos de Excelencia FB 0807,  Concurso Inserci\'{o}n en la Academia-Folio 791220017. In Brazil, we acknowledge support from FAPERJ and CNPq.

\bibliographystyle{unsrt}
\bibliography{main}

\appendix
\section{Appendix A: Segments with identical magnetic volume (TWT0 and WTW0)}

It has been shown in the main text that one can tailor the level of the intermediate plateau in the hysteresis by controlling the length of the tube and wire segments.
Finally, we show that it is possible to get an intermediate plateau at $M/M_s \approx 0$ by choosing the length of the wire and tube segments in such a way that the total magnetic volume of the wire section(s) equal that of the tube section(s).
The internal and external diameters are identical to those of TWT4 and WTW4 (see Table~\ref{tab:parameters}).

\begin{figure*}[!ht]
\centering
\begin{subfigure}[t]{0.5\textwidth}
    \includegraphics[width=\textwidth]{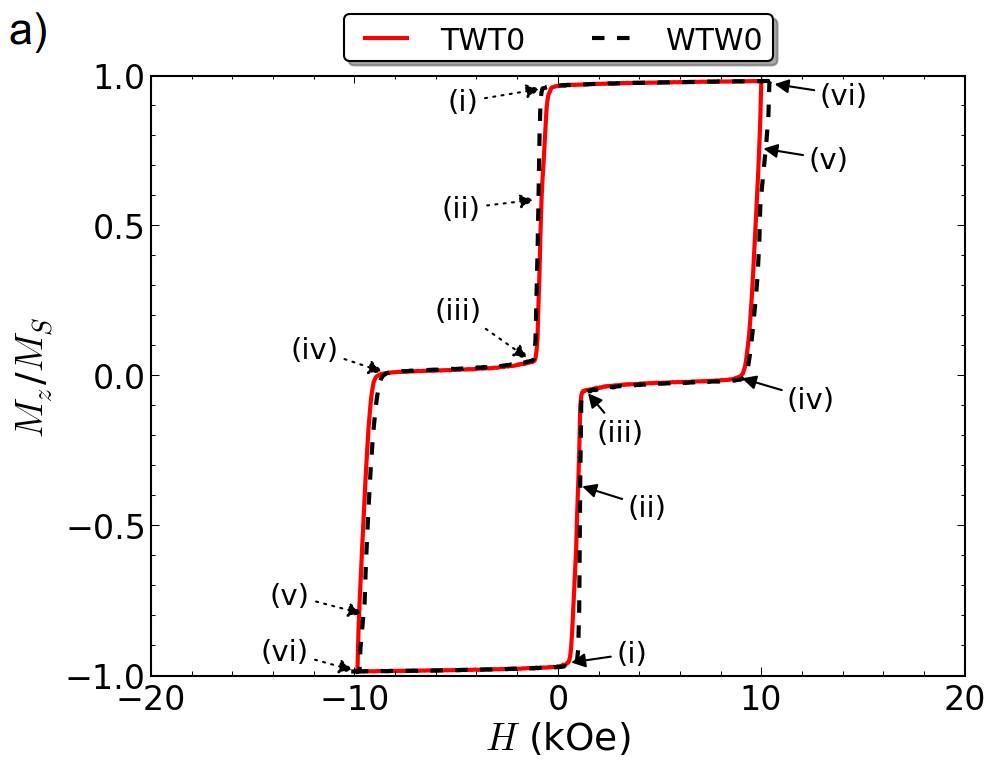}
    \caption{Hysteresis curves}
    \label{fig:TWT0-WTW0_hyst}
\end{subfigure}
\begin{subfigure}[t]{0.5\textwidth}
    \includegraphics[width=\textwidth]{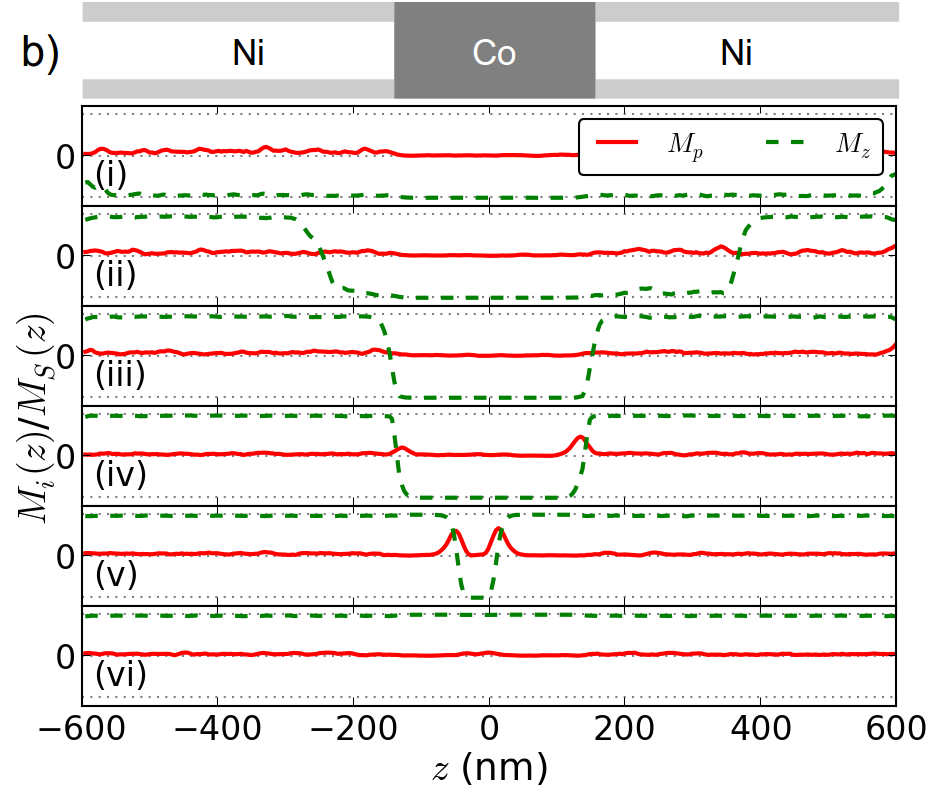}
    \caption{Magnetization profile for TWT0}
    \label{fig:TWT0_prof}
\end{subfigure}
\begin{subfigure}[t]{0.5\textwidth}
    \includegraphics[width=\textwidth]{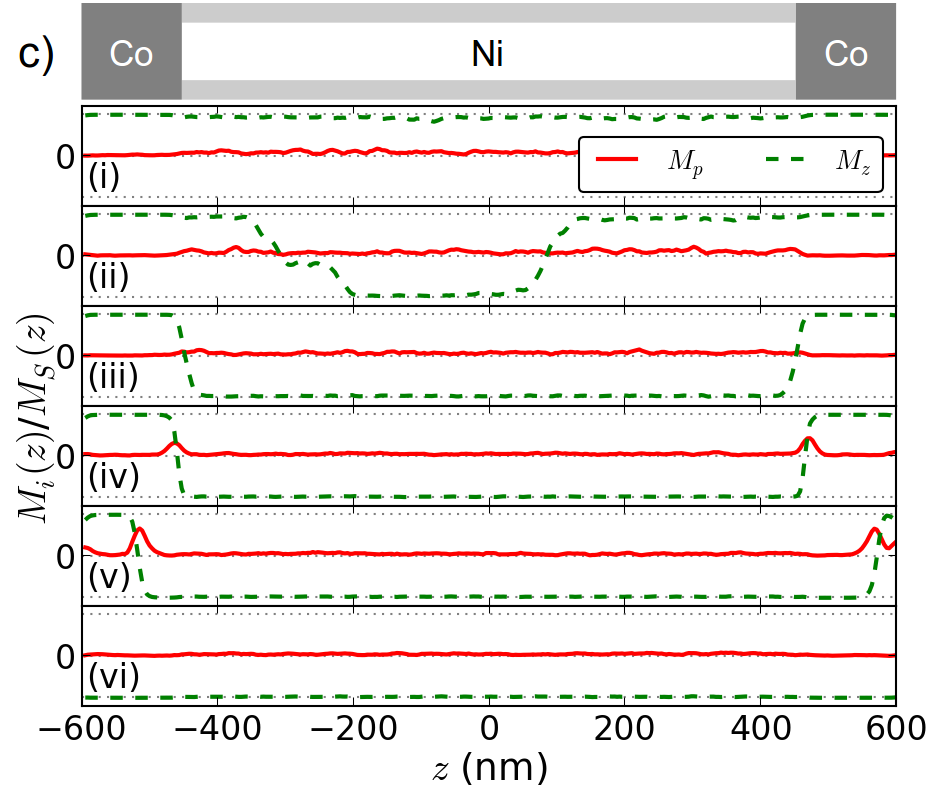}
    \caption{Magnetization profile for WTW0}
    \label{fig:WTW0_prof}
\end{subfigure}
\caption{Reversal process at $T = 300~\mbox{K}$ for elements TWT0 and WTW0, in which the total magnetic volume of wire and tube sections was chosen to be the same.
(\subref{fig:TWT0-WTW0_hyst}) Hysteresis curves for an applied field along the axis. The solid and dotted arrows indicate the points for which the magnetization profile is plotted for TWT0 and WTW0, respectively.
Magnetization profile for selected points during the hysteresis of (\subref{fig:TWT0_prof}) TWT0 and (\subref{fig:WTW0_prof}) WTW0.}
\label{fig:elements_0}
\end{figure*}

The corresponding simulations are shown in Figure~\ref{fig:elements_0}.
It should be noted that the hysteresis curves coincide for both elements.
It is clear, therefore, that the hysteresis curve alone, such as Figure~\ref{fig:TWT0-WTW0_hyst}, cannot tell these two systems apart.
It requires one to analyse the magnetization profile during reversal, as in Figures~\ref{fig:TWT0_prof} and~\ref{fig:WTW0_prof}, to fully understand the behaviour of the systems.

\end{document}